\documentclass[aps,superscriptaddress,twocolumn,floatfix]{revtex4}

\usepackage{amssymb}
\usepackage{amsmath}
\usepackage{graphicx}
\usepackage{verbatim}
\usepackage[normalem]{ulem}
\usepackage[dvips]{color}
\usepackage{multirow}
\usepackage{dcolumn}                 
\usepackage[bookmarksnumbered,bookmarksopen,colorlinks,citecolor=blue,linkcolor=blue]{hyperref} %
\usepackage{CJK}                     
\usepackage[normalem]{ulem} 
\usepackage[dvips]{color} 
\renewcommand\sout{\bgroup \color{red} \ULdepth=-.5ex \ULset}

\usepackage{epstopdf, color}

\def\esym{$E_{\rm{sym}}(\rho)$~}
\def\rpi {$\pi^-/\pi^+$~}
\def\es0{$E_{sym}(\rho_0)$~}
\begin{document}
\begin{CJK*}{GBK}{song}
\title{Effects of retarded electrical fields on observables sensitive to the high-density behavior of nuclear symmetry energy in heavy-ion collisions at intermediate energies}
\author{Gao-Feng Wei}\email[Corresponding author. E-mail: ]{wei.gaofeng@foxmail.com}
\affiliation{School of Mechanical and Material Engineering, Xi'an University of Arts and Science, Xi'an 710065, China}
\author{Bao-An Li}
\affiliation{Department of Physics and Astronomy, Texas A$\&$M
University-Commerce, Commerce, TX 75429-3011, USA}
\author{Gao-Chan Yong}
\affiliation{Institute of Modern Physics,
Chinese Academy of Sciences, Lanzhou 730000, China}
\author{Li Ou}
\affiliation{College of Physics and Technology, Guangxi Normal University, Guilin 541004, China}
\affiliation{Guangxi Key Laboratory Breeding Base of Nuclear Physics and Technology, Guilin 541004, China}
\author{Xin-Wei Cao}
\affiliation{School of Mechanical and Material Engineering, Xi'an University of Arts and Science, Xi'an 710065, China}
\author{Xu-Yang Liu}
\affiliation{School of Mathematics and Physics, Bohai University, Jinzhou 121013, China}

\begin{abstract}
Within the isospin- and momentum-dependent transport model IBUU11, we examine the relativistic retardation effects of electrical fields on the \rpi ratio and neutron-proton differential transverse flow in heavy-ion collisions at intermediate energies. Compared to the static Coulomb fields, the retarded electric fields of fast-moving charges are known to be anisotropic and the associated relativistic corrections can be significant. They are found to increase the number of energetic protons in the participant region at the maximum compression by as much as 25\% but that of energetic neutrons by less than 10\% in $^{197}$Au+$^{197}$Au reactions at a beam energy of 400 MeV/nucleon. Consequently, more $\pi^{+}$ and relatively less $\pi^{-}$ mesons are produced, leading to an appreciable reduction of the \rpi ratio compared to calculations with the static Coulomb fields. Also, the neutron-proton differential transverse flow, as another sensitive probe of high-density symmetry energy, is also decreased appreciably due to the stronger retarded electrical fields in directions perpendicular to the velocities of fast-moving charges compared to calculations using the isotropic static electrical fields. Moreover, the retardation effects on these observables are found to be approximately independent of the reaction impact parameter.
\end{abstract}


\maketitle

\section{Introduction}\label{introduction}
Nuclear symmetry energy $E_{sym}(\rho)$ at supra-saturation densities is currently the most uncertain part of the Equation of State (EOS) of dense neutron-rich nucleonic matter that can be found in central heavy-ion collisions with rare isotopes in terrestrial laboratories, interiors of neutron stars and overlapping regions of cosmic collisions involving neutron stars and/or black holes. Much efforts have been devoted to extracting information about the $E_{sym}(\rho)$ from experimental observables and astrophysical messengers, see, e.g., refs.~\cite{Steiner05,ditoro,LCK08,lynch09,DiToro10,Lat13,Trau12,Tsang12,Hor14,LiBA14,Heb15,Bal16,Wolfgang16,Ran16,Oer17,LiNews} for comprehensive reviews. Central heavy-ion reactions play a special role in this endeavor as they are the unique tools available in terrestrial laboratories to form dense neutron-rich matter. The heavy-ion reaction community has actually identified several promising probes of the high-density behavior of $E_{sym}(\rho)$. In particular, the ratio \rpi of charged pions and the neutron-proton differential transverse flow
have been found consistently using several transport models, e.g., refs \cite{Bao00,Yong06,Xie13,AMD,Cozma16,Tsang17}, to be among observables most sensitive to the high-density behavior of nuclear symmetry energy. For example, these different models all agree qualitatively that a larger (lower) value of $E_{sym}(\rho)$ at supra-saturation densities will lead to a more neutron-poor (rich) participant region and subsequently a lower (higher) \rpi ratio. Quantitatively, however, it has also been shown in several studies that current predictions are still too model and interaction dependent \cite{Li02NPA,Ditoro05,Xiao09,Feng10,XuKo10,guo13,Hong14,Song15,Wei16} to make a strong conclusion about the high-density $E_{sym}(\rho)$ from comparing calculations with existing \rpi data \cite{FOPI}.  While new experiments are being carried out \cite{ASY-EOS,Shane}, more theoretical efforts have been devoted recently to investigating various uncertain aspects associated with pion production in heavy-ion collisions. These include the in-medium pion potential \cite{XuKo10,Hong14,Song15,Guo15a,Zhang17,Feng17}, the isovector potential of $\Delta$(1232) resonances \cite{Uma98,Bao15a,Guo15b}, neutron-skins of colliding nuclei \cite{Wei14}, and tensor-force-induced short-range correlations \cite{Bao15b,Yong16}. Moreover, the transport reaction theory communities have been making efforts to compare codes to better understand model dependence, identify best practices and develop new strategies to more reliably extract information about the high-density symmetry energy from heavy-ion reactions at intermediate energies \cite{kolo05,trans1,trans2}.

It is well known that the Coulomb field affects significantly the spectrum ratio of charged pions in heavy-ion reactions, see, e.g., refs. \cite{Bertsch80,Li95,OSA96,NA44,Teis,Gor97,Fuchs98,Wagner98,Barz98,Ryb07}. To our best knowledge, in most of dynamical simulations of heavy-ion reactions so far only the standard Coulomb field $E_{static}$ is used. However, it is also well known that the first-order relativistic correction to the electric field created by a moving charge of velocity $\vec{v}$ is $E_{static}\cdot [1+(3cos(\theta)-1)\cdot v/c)]$ where $\theta$ is the angle between $\vec{v}$ and the field position vector according to the Li\'{e}nard-Wiechert formula. The correction is angular dependent and significant for fast-moving particles. For the field points along the direction of motion of the charged particle, the correction is $2\cdot v/c$ which may have significant effects on charged pions or even protons in heavy-ion collisions at intermediate energies.
Moreover, the retarded electric field is the strongest in the direction perpendicular to the velocity of the charged particle instead of being isotropic as the static Coulomb field. Thus, it is useful to examine how the relativistically retarded electrical field may affect some experimental observables known to be sensitive to the high-density behavior of nuclear symmetry energy.

In this work, effects of relativistically retarded electric fields on the \rpi ratio and neutron-proton differential transverse flow are studied in heavy-ion reactions at intermediate energies. We found that the retarded electric fields increase the number of energetic protons (neutrons) in the participant region by as much as 25\% (less than 10 \% as a secondary effect) leading to relatively more $\pi^+$ production, thus a reduction of the \rpi ratio by about 8\% compared with calculations using the static Coulomb field as normally done in transport model simulations of heavy-ion collisions at intermediate energies. Appreciable effects on the neutron-proton differential transverse flow are also found. Moreover, these retardation effects are found to be approximately independent of the impact parameter of the reaction. Thus, as an intrinsic feature of electrical interactions associated with high-speed charged particles, relativistic retardation effects should be considered to predict more precisely the \rpi ratio and neutron-proton differential transverse flow in heavy-ion collisions at intermediate energies.

In the following, we first outline the major ingredients of the isospin- and momentum-dependent Boltzmann-Uehling-Uhlenbeck transport model (IBUU) \cite{IBUU} and recall the Li\'{e}nard-Wiechert formalism in Section II. We then discuss our results in Section III. A summary will be given in Section IV.

\section{The IBUU transport model incorporating Li\'{e}nard-Wiechert potentials}\label{model}
The present study is carried out within the IBUU transport model \cite{IBUU}. In the IBUU11 version of this model,
the nuclear mean-field interaction is expressed as \cite{Das03,Chen14b}
\begin{eqnarray}
U(\rho,\delta ,\vec{p},\tau ) &=&A_{u}(x)\frac{\rho _{-\tau }}{\rho _{0}}%
+A_{l}(x)\frac{\rho _{\tau }}{\rho _{0}}+\frac{B}{2}{\big(}\frac{2\rho_{\tau} }{\rho _{0}}{\big)}^{\sigma }(1-x)  \notag \\
&+&\frac{2B}{%
\sigma +1}{\big(}\frac{\rho}{\rho _{0}}{\big)}^{\sigma }(1+x)\frac{\rho_{-\tau}}{\rho}{\big[}1+(\sigma-1)\frac{\rho_{\tau}}{\rho}{\big]}
\notag \\
&+&\frac{2C_{\tau ,\tau }}{\rho _{0}}\int d^{3}p^{\prime }\frac{f_{\tau }(%
\vec{p}^{\prime })}{1+(\vec{p}-\vec{p}^{\prime })^{2}/\Lambda ^{2}}
\notag \\
&+&\frac{2C_{\tau ,-\tau }}{\rho _{0}}\int d^{3}p^{\prime }\frac{f_{-\tau }(%
\vec{p}^{\prime })}{1+(\vec{p}-\vec{p}^{\prime })^{2}/\Lambda ^{2}}.
\label{MDIU}
\end{eqnarray}%
In the above expression, $\rho=\rho_n+\rho_p$ is the nucleon number density and $\delta=(\rho_n-\rho_p)/\rho$ is the isospin asymmetry of the nuclear medium; $\rho_{n(p)}$ denotes the neutron (proton) density, the isospin $\tau$ is $1/2$ for neutrons and $-1/2$ for protons, and $f(\vec{p})$ is the local phase space distribution function. The parameters $A_{l}(x)$ and $A_{u}(x)$ are in forms of \cite{Chen14b}
\begin{eqnarray}
A_{l}(x)&=&A_{l0} - \frac{2B}{\sigma+1}\big{[}\frac{(1-x)}{4}\sigma(\sigma+1)-\frac{1+x}{2}\big{]},  \\
A_{u}(x)&=&A_{u0} + \frac{2B}{\sigma+1}\big{[}\frac{(1-x)}{4}\sigma(\sigma+1)-\frac{1+x}{2}\big{]}.
\end{eqnarray}

\begin{figure}[h]
\centerline{\includegraphics[width=1.1\columnwidth]{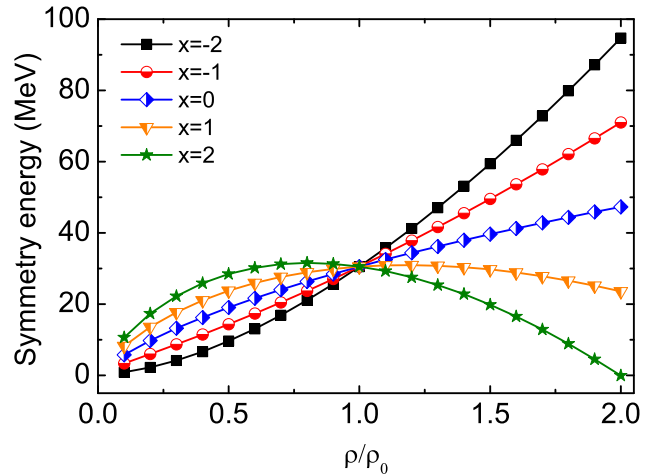}}
\caption{(Color online) The density dependence of nuclear symmetry energy $E_{sym}(\rho)$.} \label{esym}
\end{figure}

Compared to the IBUU04 version \cite{IBUU} of the model where the modified Gogny MDI (Momentum-Dependent-Interaction) is used, the adjusted parameters of $A_{l}(x)$, $A_{u}(x)$, $C_{\tau ,\tau }$, $C_{\tau ,-\tau }$ used in IBUU11 take into account more accurately the spin-isospin dependence of in-medium effective many-body forces by distinguishing the density dependences of nn, pp and np interactions in the effective 3-body force term \cite{CXu10}. They can also better fit the high-momentum behaviors of both the isoscalar and isovector nucleon optical potential extracted from nucleon-nucleus scattering experiments \cite{LXH13}.
Using empirical constraints and properties of symmetric nuclear matter at normal density, the values of these parameters can be determined as $A_{l0}(x)$ = -76.963 MeV, $A_{u0}(x)$ = - 56.963 MeV, $B$= 141.963 MeV, $C_{\tau ,\tau }$= -57.209 MeV, $C_{\tau ,-\tau }$= -102.979 MeV, $\sigma$= 1.2652 and $\Lambda $= 2.424$p_{f0}$ where $p_{f0}$ is the nucleon Fermi momentum in symmetric nuclear matter at normal density. They lead to a binding energy of $-16$ MeV, an incompressibility of $230$ MeV for symmetric nuclear matter and a symmetry energy $E_{sym}(\rho_0)=30.0$ MeV at the saturation density of $\rho_0=0.16$ fm$^{-3}$. The parameter $x$ is introduced to mimic the different forms of symmetry energy predicted by various many-body theories without changing any property of symmetric nuclear matter and the value of symmetry energy at saturation density. The density dependences of nuclear symmetry energy with different $x$ parameters are shown in Fig. \ref{esym}.

In one of our previous works \cite{Ou11}, the complete Li\'{e}nard-Wiechert potentials for both electrical and magnetic fields are consistently incorporated into the IBUU11 code. As it was shown both analytically and numerically in detail in ref. \cite{Ou11}, the ratio of Lorentz force over the Coulomb force is approximately $(v/c)^2$. The Lorentz force was found to have negligible effect on the ratio of charged pions except at extremely forward/backward rapidities.
Thus, unless we consider the second-order relativistic correction to the electrical fields, one can safely neglect the Lorentz force and thus speed up the code dramatically by turning off the calculation of the magnetic fields. In this work, thus only the electric fields are calculated according to the well-known Li\'{e}nard-Wiechert expression
\begin{equation}\label{electrical field}
e\vec{E}(\vec{r},t)=\frac{e^2}{4\pi \varepsilon_{0}}
\sum_{n}Z_{n}\frac{c^2-v^{2}_{n}}{(cR_{n}-\vec{R}_{n}\cdot \vec{v}_{n})^3}(c\vec{R}_{n}-R_{n}\vec{v}_{n})
\end{equation}
where $Z_{n}$ is the charge number of the $n$th particle, $\vec{R}_{n}=\vec{r}-\vec{r}_{n}$ is the position of the field point $\vec{r}$ relative to the source
point $\vec{r}_{n}$ where the $n$th particle is moving with velocity $\vec{v}_{n}$ at the retarded time of $t_{n}=t-|\vec{r}-\vec{r}_{n}|/c$.
Naturally, in the nonrelativistic limit $v_{n}$$\ll$$c$,
Eq.(\ref{electrical field}) reduces to the static Coulomb field of the form
\begin{equation}\label{Coulomb field}
e\vec{E}(\vec{r},t)=\frac{e^2}{4\pi \varepsilon_{0}}
\sum_{n}Z_{n}\frac{\vec{R}_{n}}{R_{n}^{3}}.
\end{equation}
Obviously, the most important differences between the two formulas in Eqs. (\ref{electrical field} and \ref{Coulomb field}) are the relativistic retardation effects and the non-isotropic nature of the retarded electrical fields of fast-moving charges. In the relativistic case, all charged particles with velocity $\vec{v}_{n}$ at the retarded time $t_n$ contribute to the $e\vec{E}(\vec{r},t)$ at the instant $t$ and location $\vec{r}$. Whereas in the nonrelativistic case, those charged particles contribute to the $e\vec{E}(\vec{r},t)$
only at the same moment $t$. Of course, the retardation effect depends on the reduced velocity $\beta=v/c$. In a typical reaction at a beam energy of 400 MeV/nucleon available in several laboratories, the velocities of charged protons reach about $0.7c$. It is high enough to warrant an investigation about effects of the retarded electrical fields on several observables useful for studying the high-density behavior of nuclear symmetry energy. It is well known that for a charge moving with a constant velocity $\vec{v}$ with respect to the rest frame S, its electrical field seen by an observer at rest in S is asymmetric, i.e., longitudinally reduced while transversely enhanced such that it may look like a pancake at ultra-relativistic energies. More quantitatively, while the electrical field is enhanced by a factor $\gamma=1/(1-\beta^2)^{1/2}$ perpendicular to $\vec{v}$, it is weakened in the direction of motion by $1/\gamma^2$. In heavy-ion collisions at intermediate and higher beam energies, the two Lorentz-contracted nuclei (both in terms of their electrical fields and matter distributions) moving in the opposite directions come to collide with each other under the influence of both strong and Coulomb forces as well as frequent nucleon-nucleon collisions. The complicated dynamics of such reactions is simulated by using the IBUU11 transport model in this work.

\begin{figure}[th]
\centerline{\includegraphics[width=1.1\columnwidth]{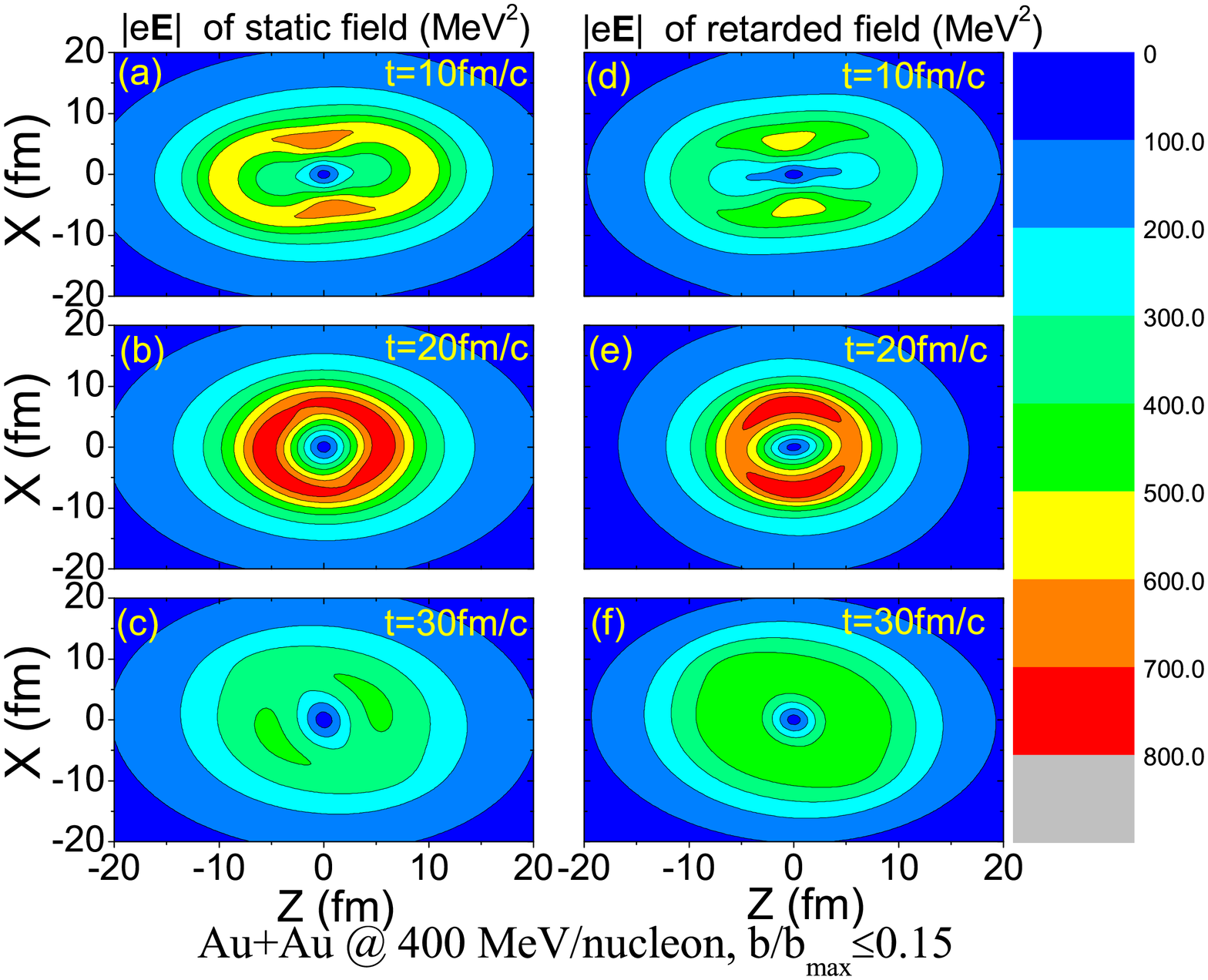}}
\vspace{0.5cm}
\centerline{\includegraphics[width=1.1\columnwidth]{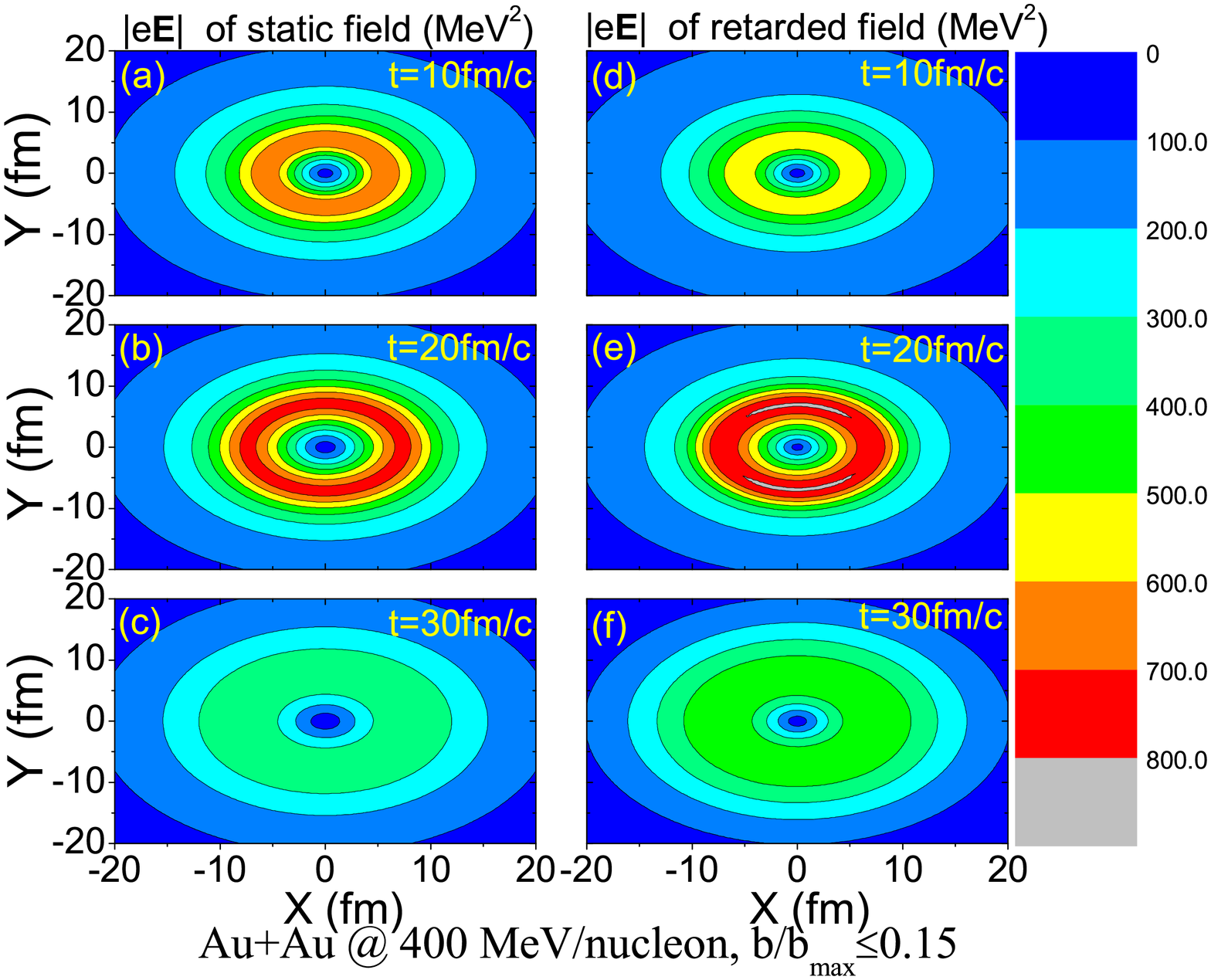}}
\caption{(Color online) Contours of the electric fields in the X-o-Z reaction plane (upper) and X-o-Y plane (lower) at the initial compression (t=10 fm/$c$), maximum compression (t=20 fm/$c$) and expansion stage (t=30 fm/$c$) in central Au+Au collisions at a beam energy of 400MeV/nucleon, respectively. The panels (a), (b) and (c) are for the static fields while the panels (d), (e) and (f) are for the retarded
fields. } \label{fld}
\end{figure}

\section{Results and Discussions}\label{results}
First of all, it is necessary to mention that to calculate the retarded electric fields $e\vec{E}(\vec{r},t)$, the phase space histories of all charged particles before the moment $t$ have to be saved in transport model simulations. Moreover,
a pre-collision phase space history for all nucleons is made assuming that they are frozen in the projectile and target moving along their Coulomb trajectories. More technical details about calculating the $e\vec{E}(\vec{r},t)$ and numerical checks in idealized cases against analytical solutions can be found in ref. \cite{Ou11}. In the following illustrations, we present results for the reaction of $^{197}$Au+$^{197}$Au at a beam energy of 400 MeV/nucleon. The calculations are done in the CMS frame of the colliding nuclei. The beam is in the Z-direction and the reaction plane is the X-o-Z plane. In the following discussions, we refer the electrical fields calculated using the Li\'{e}nard-Wiechert formula as the retarded fields while those from using the normal Coulomb formula in each time step as the static electrical fields. But we do generally refer all electrical forces as the Coulomb force.
\subsection{Evolutions of the anisotropic retarded electric fields in comparison with the static Coulomb fields}
To help understand effects of the retarded electric fields on the reaction dynamics and experimental observables sensitive to the symmetry energy in heavy-ion collisions, we show and discuss in this subsection the time evolution and space distribution of retarded electric fields in comparison with the static Coulomb fields normally used in simulating heavy-ion reactions. We present here results for central reactions with impact parameters $b/b_{max}\leq 0.15$.

Shown in the upper window of Fig. \ref{fld} are the strength $|e\bold{E}|$ contours of the static and retarded electric fields in the reaction plane (X-o-Z) at three instants representing the initial compression, maximum compression and the expanding stage, respectively. Firstly, at the initial compression stage, both the static and retarded electric fields show two strong regions around the centers of the target and projectile above and below the origin of the coordinate. When the two electrical pancakes of projectile and target passing each other, the total electrical field is zero around the origin of the coordinate. Secondly, the retarded electric field is weaker (stronger) than the static electric field at the initial compression (expanding) stage as one expects due to the time delay in the relativistic calculations. Thirdly, at the maximum compression when the system has been sufficiently stoped and thermalized, the retarded electric field is obviously anisotropic in the reaction plane compared with the isotropic static electric field. The fields are appreciably stronger in the X direction but weaker in the Z direction. This feature is qualitatively what the Li\'{e}nard-Wiechert formula predicts for moving charges. Of course, as we discussed earlier the dynamics of heavy-ion collisions are much more complicated than two approaching charges.
\begin{figure}[h]
\centerline{\includegraphics[width=1.1\columnwidth]{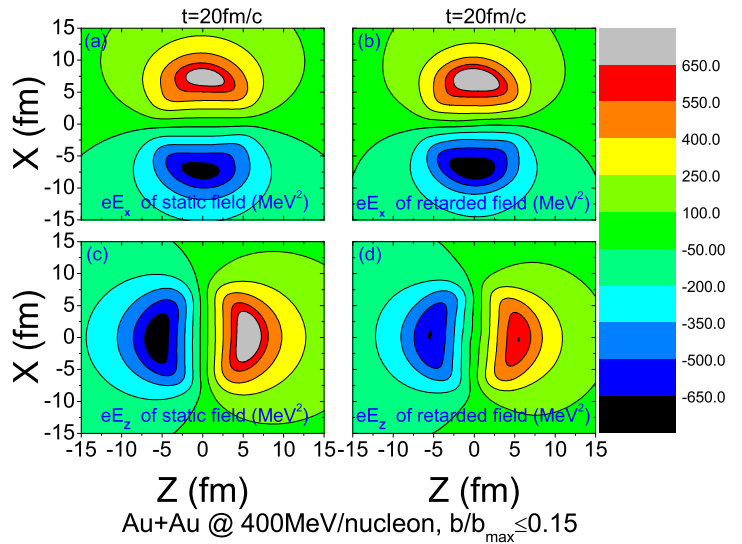}}
\vspace{0.5cm}
\centerline{\includegraphics[width=1.1\columnwidth]{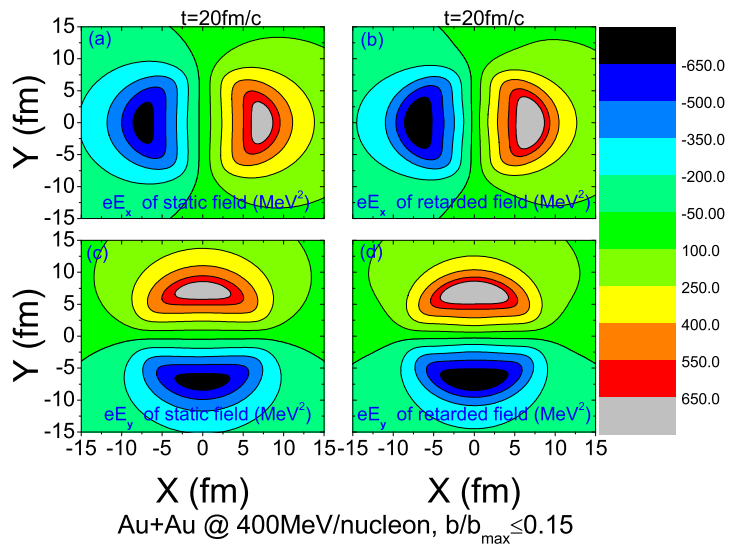}}
\caption{(Color online) Upper: Contours of the X and Z components of the electric fields in the X-o-Z plane at the maximum compression
(t=20 fm/$c$) in central Au+Au collisions at a beam energy of 400MeV/nucleon, respectively.
Lower: Contours of the X and Y components of the electric fields in the X-o-Y plane.
The panels (a) and (c) are for the static fields while the panels (b) and (d) are for the retarded fields. } \label{comfld}
\end{figure}
Because of symmetries in the plane perpendicular to the beam direction, the strengths $|e\bold{E}|$ of the electrical fields in the X-o-Y plane are approximately spherically symmetric as shown in the lower window of Fig. \ref{fld}. While overall the nuclei are moving in the $\pm Z$ direction, because of the Fermi motion in the initial state, nucleon-nucleon collisions, before reaching the maximum compressions, some particles have obtained significant velocity components in the X and Y directions although they are still less than the velocity component in the Z direction before a complete thermalization is realized. Thus, at the stage of maximum compression the electrical fields are still asymmetric in the X-o-Z plane. In more detail and quantitatively, shown in the upper window of Fig. \ref{comfld} are comparisons of the $eE_x$ and $eE_z$ in the reaction plane at the instant of 20 fm/c. For the retarded fields, vertically it is seen that the $eE_x$ in the $\pm X$ directions is significantly higher and covers a larger area than the  $eE_z$ in the $\pm Z$ direction. While for the static fields the $eE_x$ and $eE_z$ are very close to each other as shown in the left panels of Fig. \ref{comfld}.  In the X-o-Y plane, however, as shown in the lower window of Fig. \ref{comfld} the $eE_x$ and $eE_y$ are very close to each other for both the retarded and static fields.

\begin{figure}[th]
\centerline{\includegraphics[width=1.1\columnwidth]{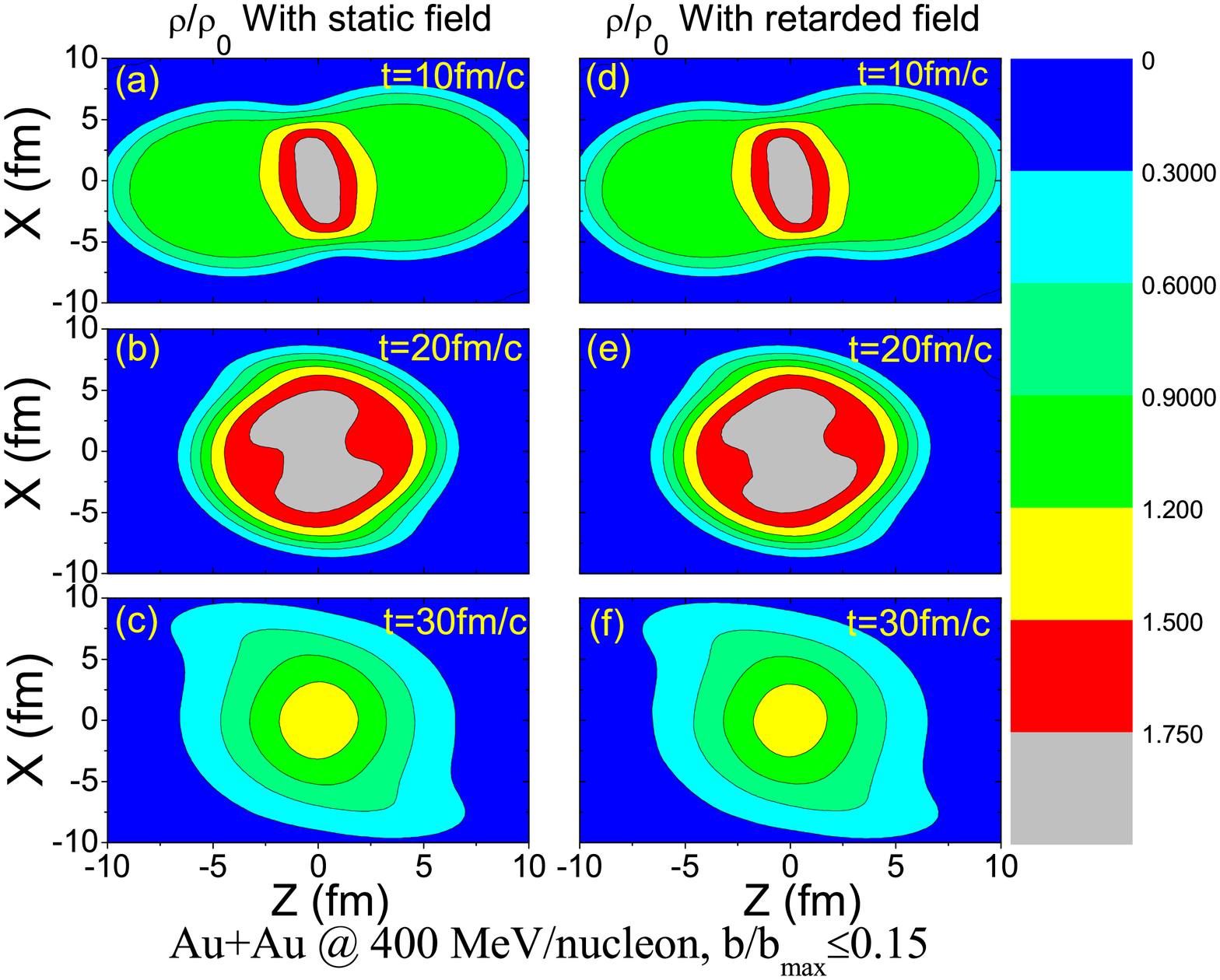}}
\vspace{0.5cm}
\centerline{\includegraphics[width=1.1\columnwidth]{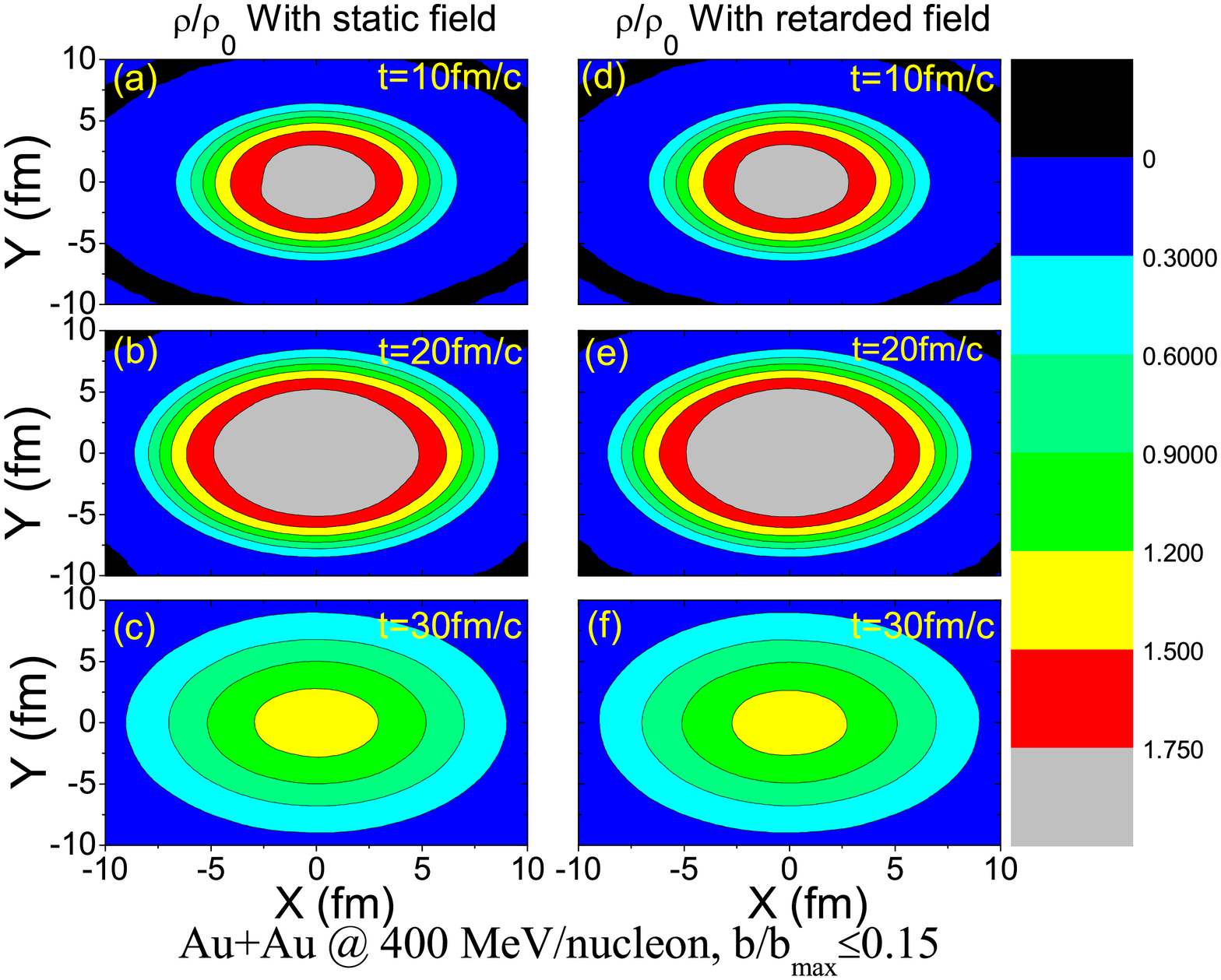}}
\caption{(Color online) Density contours in the X-o-Z plane (upper) and X-o-Y plane (lower) at the initial compression (t=10 fm/$c$),
maximum compression (t=20 fm/$c$) and expansion stage (t=30 fm/$c$) in central Au+Au collisions at a beam energy of 400 MeV/nucleon, respectively.
The panels (a), (b) and (c) are for the static fields while the panels (d), (e) and (f) are for the retarded
fields. } \label{den}
\end{figure}

Because the Coulomb force is much smaller than the nuclear force, we do not expect the overall dynamics and global properties of nuclear reactions are affected by whether one uses the static or retarded electrical fields. Indeed, as shown in the density contours in the X-o-Z and X-o-Y planes in Fig. \ref{den}, respectively, their evolutions and distributions are very similar. However, it is worth noting that at 20 fm/c, as shown in the middle of the upper-right window of Fig. \ref{den} the retarded electric field leads to a slightly larger high density region due to the weakened repulsive force in the Z direction compared to the static calculation as we discussed above.
\subsection{Effects of the retarded electrical fields on the \rpi ratio}
\begin{figure}[th]
\centerline{\includegraphics[width=1.1\columnwidth]{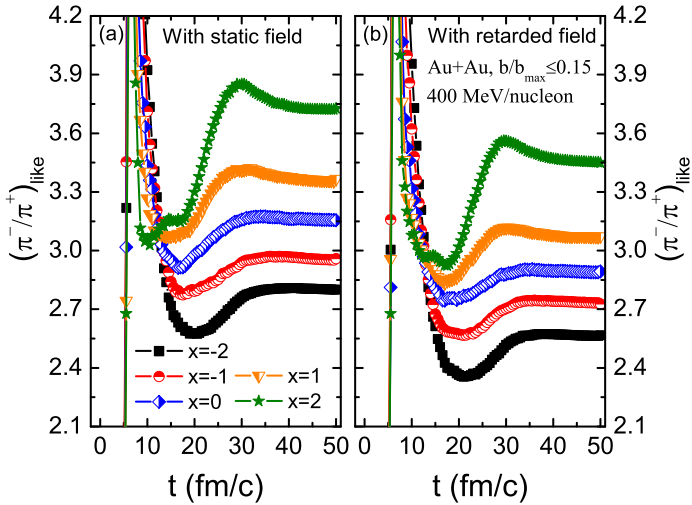}}
\caption{(Color online) Time evolution of the (\rpi)$_{\rm like}$ ratio in central
$^{197}$Au+$^{197}$Au collisions at 400MeV/nucleon with symmetry energies ranging from super-hard of $x$=-2 to super-soft of $x$=2. The panels (a) and (b) are results of calculations using the static and retarded electric fields, respectively.} \label{tpion}
\end{figure}

\begin{figure}[th]
\centerline{\includegraphics[width=1.1\columnwidth]{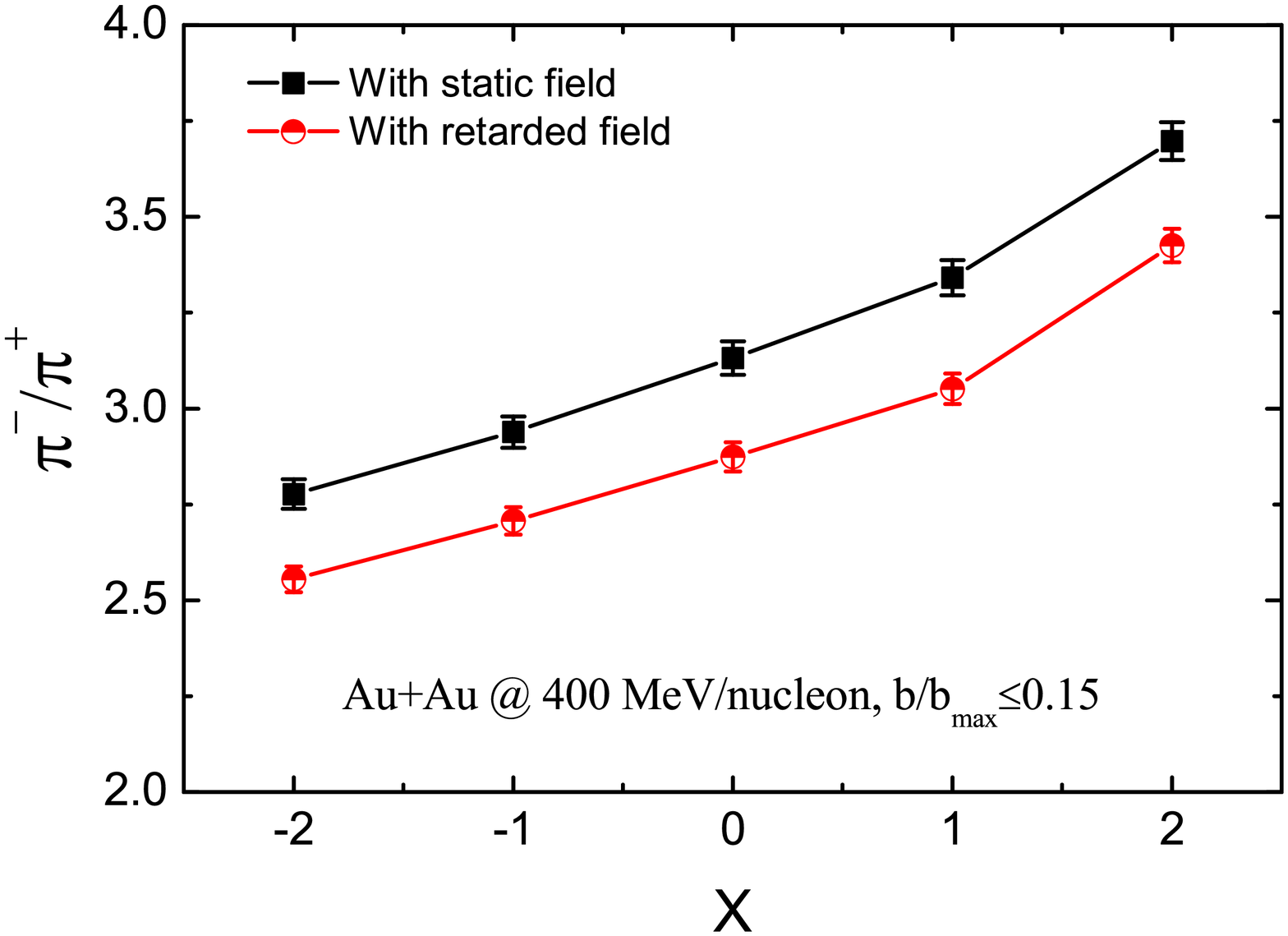}}
\centerline{\includegraphics[width=1.1\columnwidth]{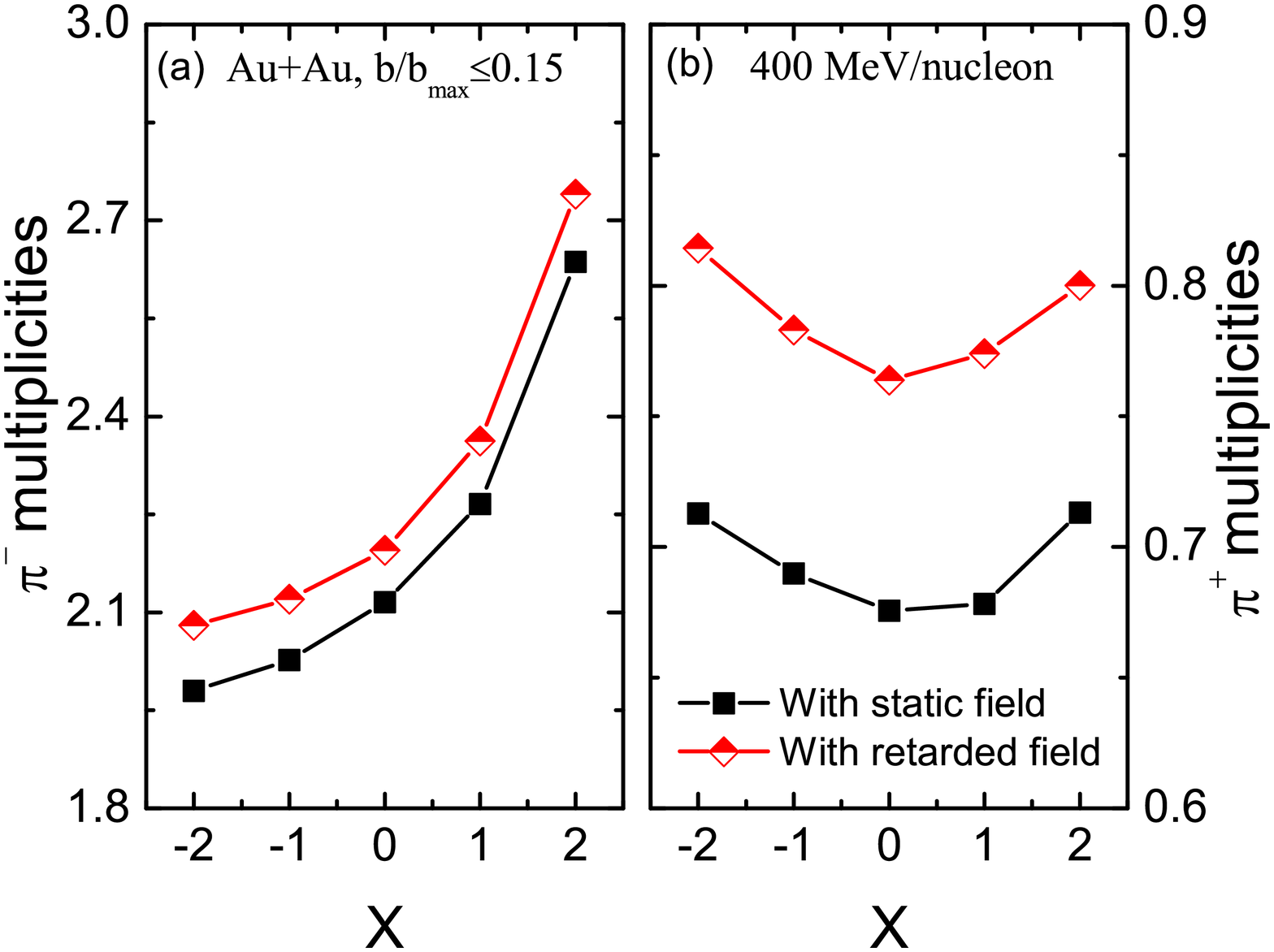}}
\caption{(Color online) Upper: The final \rpi ratio versus the symmetry energy parameter $x$ for the results shown in Fig. \ref{tpion}.
Lower: Multiplicities of $\pi^{-}$ (a) and $\pi^{+}$ (b) in the same reactions.} \label{com}
\end{figure}

\begin{figure}[th]
\centerline{\includegraphics[width=1.1\columnwidth]{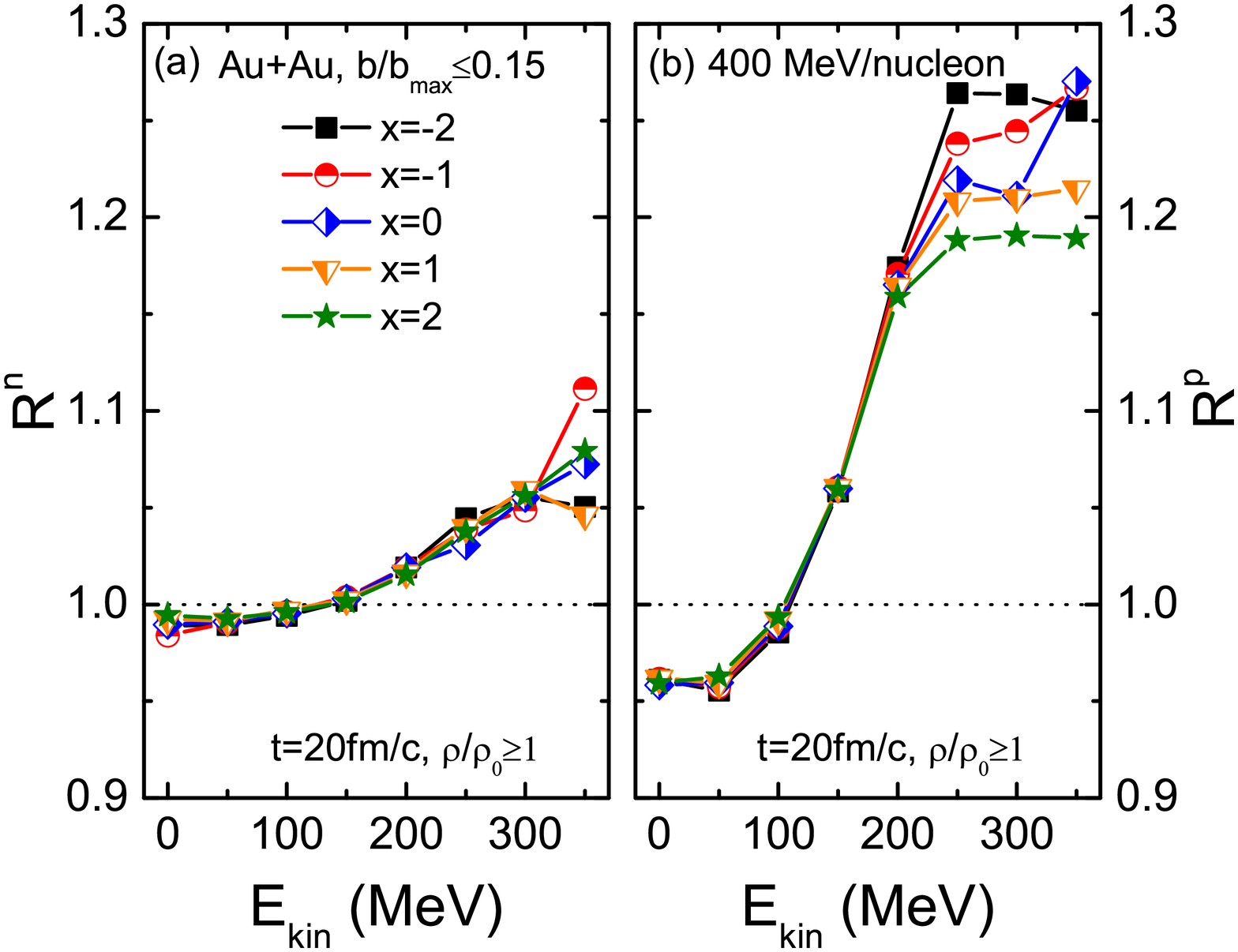}}
\caption{(Color online) The ratio $R^{n}$ (left) and $R^{p}$ (right) for neutrons (a) and  protons (b), respectively, at supra-saturation densities at the
maximum compression stage (t=20 fm/$c$) as a function of nucleon kinetic energy in central Au+Au collisions at a beam energy of 400MeV/nucleon, respectively.} \label{Rnp}
\end{figure}

We now turn to the relativistic retardation effects of electric fields on the \rpi ratio. In heavy-ion collisions at intermediate energies, pions are mostly produced from the decay of $\Delta$(1232) resonances. To examine the dynamics
of pion production in these reactions, one may use the dynamic pion ratio (\rpi)$_{\rm like}$ defined as \cite{Bao00}
\begin{equation}\label{ratio}
(\pi^{-}/\pi^{+})_{\rm like}\equiv
\frac{\pi^{-}+\Delta^{-}+\frac{1}{3}\Delta^{0}}
{\pi^{+}+\Delta^{++}+\frac{1}{3}\Delta^{+}}.
\end{equation}
Because all $\Delta$ resonances will eventually decay into nucleons and pions, the (\rpi)$_{\rm like}$ ratio will naturally become the free \rpi ratio at the end of the reaction. Shown in Fig. \ref{tpion}
is the time evolution of the (\rpi)$_{\rm like}$ ratio in central Au+Au collisions at a beam energy of 400 MeV/nucleon with retarded and static electric fields, respectively. The corresponding final \rpi ratio is shown in the upper window of Fig. \ref{com} as a function of the symmetry energy parameter $x$. Consistent with previous observations using most transport models, it is seen that the \rpi ratio is sensitive to the density dependence of nuclear symmetry energy $E_{sym}(\rho)$ regardless how the electrical fields are calculated. A softer $E_{sym}(\rho)$ leads to a higher \rpi ratio, reflecting a more neutron-rich participant region formed in the reaction.

It has been a major challenge for the transport model community to predict accurately the final \rpi ratio and agree within 20\%. Very often, the predicted effects on even the most sensitive observables, when the $E_{sym}(\rho)$ is modified from being soft to stiff within the known limits using the same model, are on the order 10-50\%.  This is mainly because the nucleon isovector potential is much weaker than the isoscalar potential. Of course, the exact sensitivity depends on the reaction system and conditions used. Therefore, better understanding various factors affecting appreciably the proposed probes of the high-density behavior of nuclear symmetry energy has been a major goal of many recent works. In this context, it is interesting to see in both Figs. \ref{tpion} and \ref{com} that the \rpi ratio at the final stage is about 8\% smaller in calculations with the retarded electric fields than those with the static ones approximately independent of the $x$ parameter used. Moreover, as shown in the lower window of Fig. \ref{com} the multiplicities of both $\pi^+$ and $\pi^-$ get increased by the retarded electrical fields. More quantitatively, the multiplicity of $\pi^+$ is increased by about 14\% while that of $\pi^-$ by less than 5\%. These results are surprising as one normally expects the Coulomb field mainly affects the spectrum ratio of charged pions but not much their individual multiplicities. It is also surprising to see that there is a small increase in the multiplicity of $\pi^-$ which is mainly from nn inelastic scatterings that are not directly affected by the variation of the electrical fields.

To understand the above observations, we investigate the relative change in nucleon kinetic energy distributions due to using the retarded electrical fields compared to the static ones.
For this purpose, we examine the ratio
\begin{equation}
R^{i}=\frac{{\rm Number}(i)_{\rm{R}}}{{\rm Number}(i)_{\rm{S}}},~~~i\equiv{\rm{neutron~or~proton}}
\end{equation}\label{R}\\
of nucleons with local densities higher than $\rho_0$ at the maximum compression stage (20 fm/c) in the Au+Au reactions with the retarded (R) and static (S) electrical fields.
Shown in Fig. \ref{Rnp} are the $R^{n}$ and $R^{p}$ as a function of nucleon kinetic energy. Interestingly, it is seen clearly that both the $R^{n}$ and $R^{p}$ are larger than 1 for energetic nucleons above about 120 MeV, indicating that the retarded electric fields indeed increase (decrease) the number of high (low) energy nucleons, especially protons. More quantitatively, the number of energetic protons (neutrons) is increased by as much as 25\% (10\%). As the system approaches the maximum compression where the thermalization is the highest, more energetic particles are being shifted continuously to lower energies. Thus, before reaching the maximum compression there are even more energetic nucleons than indicated by Fig. \ref{Rnp} with the retarded electrical fields. These increased numbers of energetic nucleons are responsible for the increased production of pions. As we discussed earlier, one of the major features of the retarded electrical field is its asymmetries, namely its longitudinal component is reduced by $1/\gamma^2$ while its transverse component gets enhanced by $\gamma$ compared to the static fields. The stronger transverse electrical field can accelerate more charged particles to higher energies. Some protons can gain enough kinetic energies to bring certain pp collisions above the pion production threshold, leading to more $\pi^+$ mesons. While neutrons are not affected directly by the electrical fields, secondary collisions between neutrons and energetic protons can increase the kinetic energies of neutrons. In addition, neutrons couple to charged $\Delta$ resonances through $\Delta^{-}\leftrightarrow n+\pi^-$ and $\Delta^{+}\leftrightarrow n+\pi^+$ reaction channels which are affected directly by the electrical fields. Thus, the kinetic energy of neutrons, consequently the $\pi^-$ multiplicity, can also be increased by the retarded electrical fields albeit at a lower level than protons and the $\pi^+$ multiplicity.

\begin{figure}[th]
\centerline{\includegraphics[width=1.1\columnwidth]{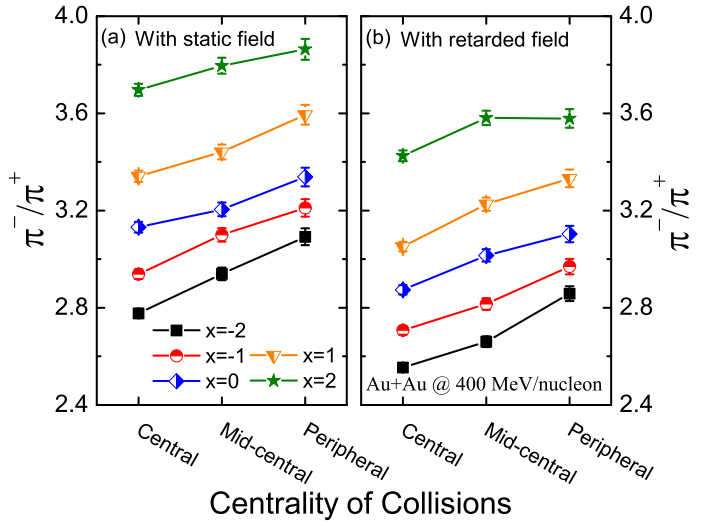}}
\caption{(Color online) The impact parameter dependence of \rpi ratio in Au+Au collisions at 400MeV/nucleon
with different symmetry energy settings from super-hard  of $x$=-2 to super-soft of $x$=2. The impact  parameters
are set as $b/b_{max}$$\leq$0.5, $b$=5 fm and $b$=7.2 fm for central, mid-central and peripheral
 collisions, respectively.} \label{impact}
\end{figure}

Next, we investigate the impact parameter dependence of relativistic retardation effects of electrical fields. Shown in Fig. \ref{impact} are the \rpi ratios in Au+Au collisions at 400 MeV/nucleon obtained using the static and retarded electric fields, respectively, as functions of centrality. It is seen that the reduction of the \rpi ratio due to the retardation effects is approximately independent of the impact parameter. Overall, since the retardation effect is an intrinsic feature of electrical interactions of charged particles moving at high speeds, given its appreciable effects on the \rpi ratio shown above, it should be considered when the \rpi ratio in heavy-ion collisions is used as a probe of high-density symmetry energy.

To this end, some discussions about comparing with available experimental data especially the ones from the FOPI collaboration \cite{FOPI} are in order. In principle, the results shown in Figs. \ref{com} and \ref{impact} can be compared to the data after considering possible detector filterings. A direct and rough comparison with the FOPI data indicates that the $\pi^+$ multiplicity is underpredicted by about 25\% approximately $x$-independently even with the retardation effect. While the $\pi^-$ multiplicity is close to the data only with the super-soft \esym of $x=2$ but is underpredicted by about 25\% with the super-stiff \esym of $x=-2$. Within the approximately 10\% uncertainty of the data, the calculated \rpi ratios with or without the retardation effect can all reasonably reproduce the data with the $x$ parameter from about 1 to -1. Since we are not considering in the present work several effects, such as the pion potential, the uncertainty of the $\Delta$ isovector potential, the isospin-dependent high-momentum tails of nucleons due to the short-range correlations in both the initial state and during the reaction, that have been shown recently to affect appreciably the \rpi ratio as we discussed in the introduction, we are unable to make a solid conclusion regarding the \esym from these comparisons. Obviously, a more comprehensive comparison of the pion data with calculations considering all of the aforementioned effects are necessary before making a final conclusion regrading the high-density behavior of nuclear symmetry energy. Nevertheless, we are confident that the relative effects of the relativistically retarded electrical fields observed in this work are physically sound and should be considered together with the other effects mentioned above in extracting eventually the \esym from analyzing the pion data.

\subsection{Effects of retarded electrical fields on neutron-proton differential transverse flow}
From Fig. \ref{Rnp} we have already seen that the retarded electrical fields affect neutrons and protons differently. Especially, the energetic nucleons are affected differently depending on the stiffness parameter $x$ of nuclear symmetry energy. Moreover, as we discussed in detail, because of the asymmetry of the retarded electrical fields the motions of charged particles are influenced differently in transverse and longitudinal directions. It is known that the neutron-proton differential transverse flow probes sensitively the nuclear isovector potential without much interference by the isoscalar potential \cite{Bao00}. Depending on the isospin asymmetry of the system, the isovector potential proportional to $\delta$ can be very small.
While the electrical force is much weaker than nuclear force, the difference between the retarded electrical field and the static one might be as big as the nuclear isovector potential even in the most neutron-rich nuclei. It is thus interesting to study if and how the relativistically retarded electrical fields may affect the neutron-proton differential transverse flow.

\begin{figure}[th]
\centerline{\includegraphics[width=1.1\columnwidth]{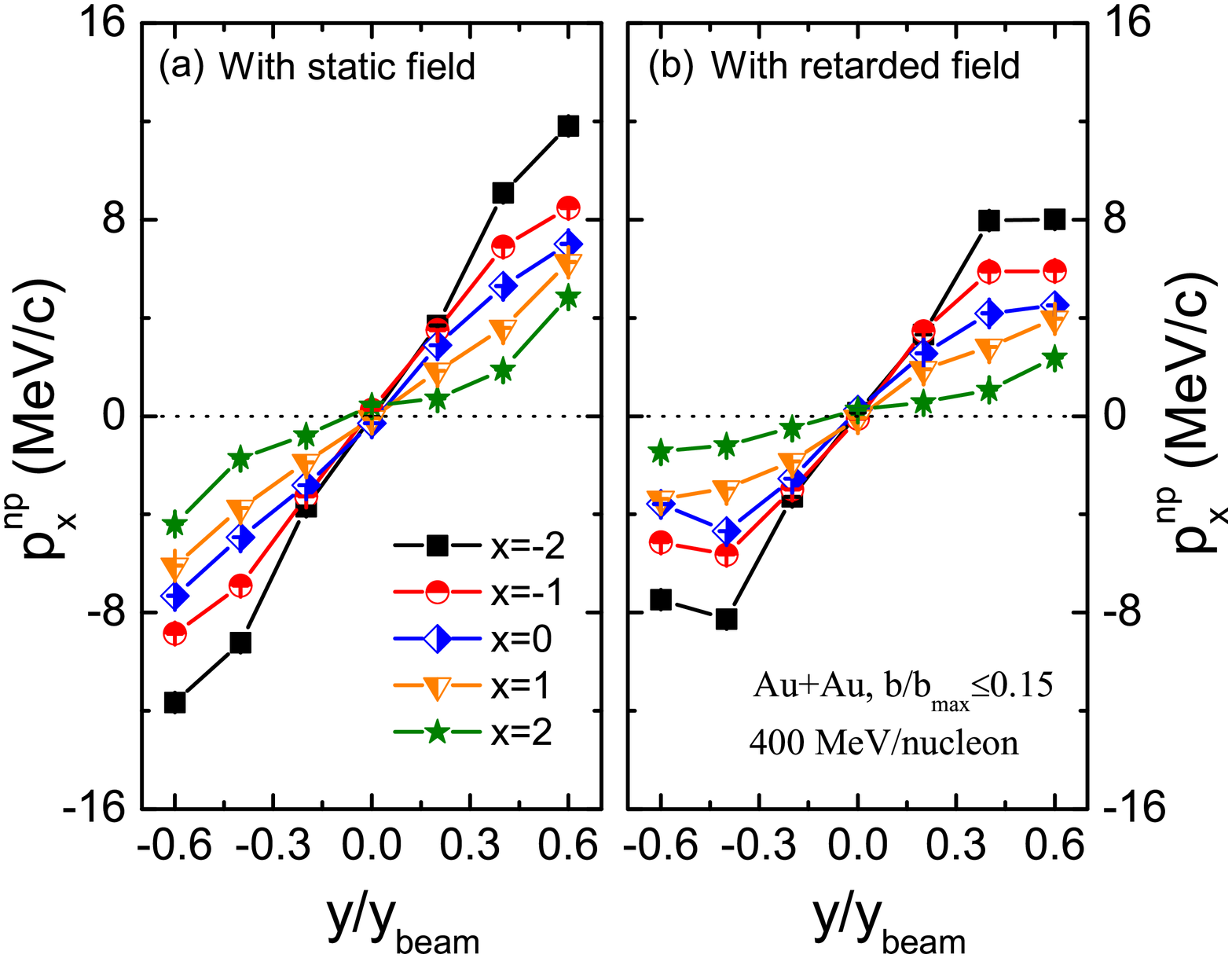}}
\vspace{0.5cm}
\centerline{\includegraphics[width=1.1\columnwidth]{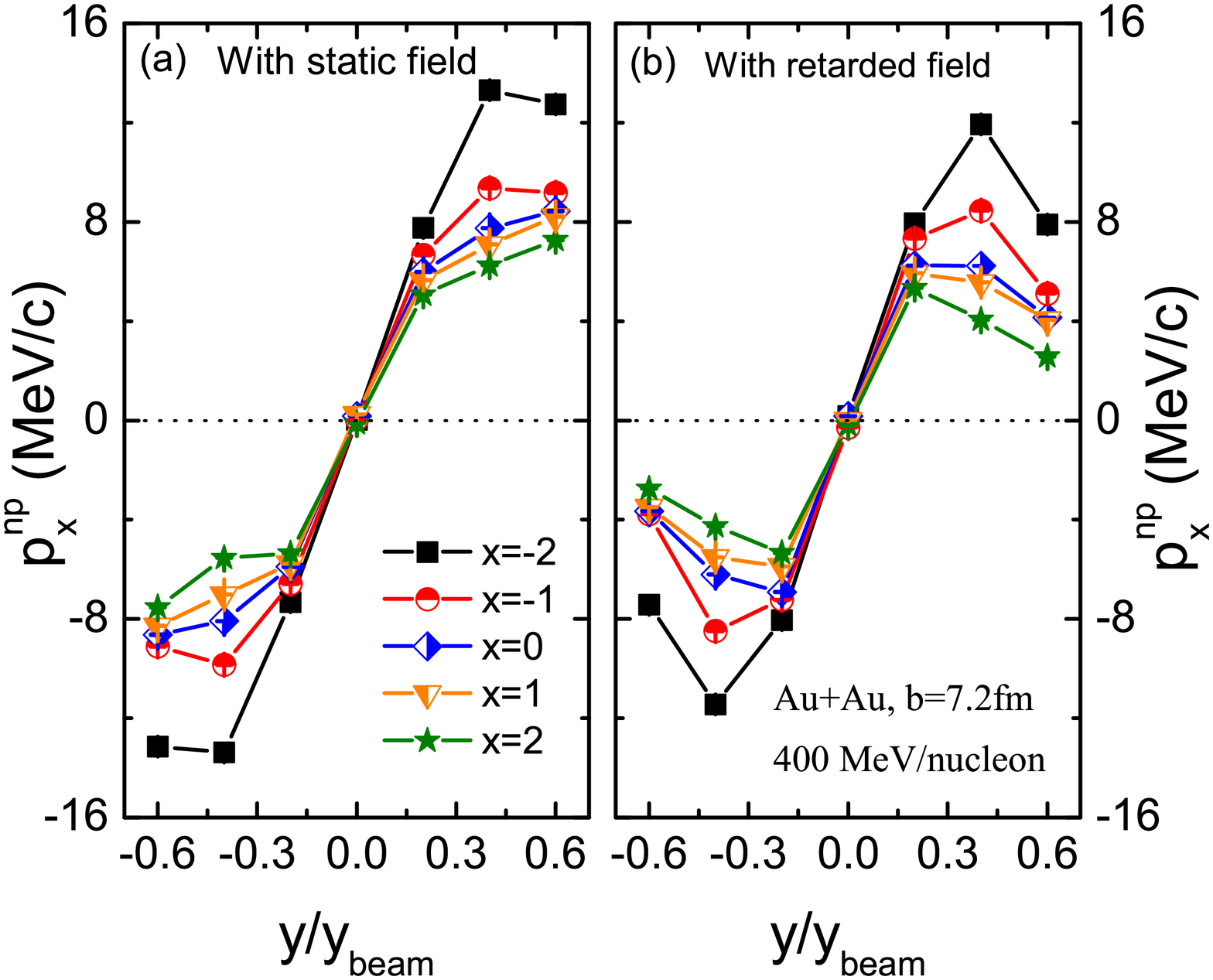}}
 \caption{(Color online) The neutron-proton differential flow in central (upper) and peripheral (lower) collisions for the same reactions and conditions as in Fig. \ref{impact}.} \label{npflow}
\end{figure}
The neutron-proton differential transverse flow was defined as \cite{Bao00}
\begin{equation}\label{np-flow}
p^{np}_{x}(y)=\frac{1}{N(y)}\sum^{N(y)}_{i=1}p_{x_{i}}\tau_{i}
\end{equation}
where $N(y)$ is the number of free nucleons with local densities less than $\rho_{0}/8$ at rapidity $y$, and $\tau_{i}$ is 1 for neutrons and -1 for protons. It was proposed as a sensitive probe of the high-density behavior of nuclear symmetry energy. It has the advantage of enhancing the signal strength of the symmetry energy by: (1) combining constructively effects of the symmetry potential on the isospin fractionation and nucleon transverse collective flow, and (2) maximizing effects of the symmetry potential while minimizing those of the isoscalar potential \cite{Bao00}. Shown in Fig. \ref{npflow} are the neutron-proton differential transverse flows in the central (upper window) and peripheral (lower window) Au+Au collisions with symmetry energies from being super-hard of $x$=-2 to super-soft of $x$=2. Obviously, the sensitivities of neutron-proton differential flow to the stiffness of symmetry energy are clearly visible irrespective of the electric fields used. Moreover, since the retarded electric fields are stronger in directions perpendicular to the velocities of charged particles, the neutron-proton differential transverse flow is reduced appreciably as a whole compared to the calculations using the isotropic static electric fields. Furthermore, the relativistic retardation effects of electric fields on the neutron-proton differential flow are approximately independent of the impact parameter of the reaction. It is worth noting here that we have also investigated effects of the retarded electrical fields on the transverse flows of neutrons and protons themselves. The effects are negligible as expected because the nuclear force overwhelms the Coulomb force.

\section{Summary}\label{summary}

In summary, we investigated effects of relativistically retarded electrical fields on the \rpi ratio and neutron-proton differential transverse flow in Au+Au collisions at a beam energy of 400 MeV/nucleon. Compared to the isotropic static Coulomb fields, the retarded electrical fields are anisotropic and strongest (weakest) in directions perpendicular (parallel) to the velocities of charged particles. As a result, some charged particles get accelerated by the enhanced electrical fields in some directions. These more energetic particles help produce more $\pi^{+}$ than $\pi^{-}$ mesons, leading to an appreciable reduction of the \rpi ratio.  Also, the neutron-proton differential transverse flow is also decreased appreciably due to the stronger retarded electrical fields in directions perpendicular to the velocities of charged particles compared to calculations using the static Coulomb fields. Moreover, these features are approximately independence of the impact parameter of the reaction. As the next step, how these effects may depend on the beam energy and N/Z ratio of the reaction system are being investigated systematically.
In conclusion, the relativistic retardation effects of electrical fields of fast-moving charges should be considered in simulating heavy-ion collisions at intermediate energies to more precisely constrain the high-density behavior of nuclear symmetry energy using the \rpi ratio and/or neutron-proton differential transverse flow as probes.

\section*{\textbf{Acknowledgements}}
G.F. Wei would like to thank Profs. Zhao-Qing Feng and Yuan Gao for helpful discussions and Prof. Shan-Gui Zhou for providing us the computing resources at the HPC Cluster
of SKLTP/ITP-CAS where some of the calculations for this work were done.
This work is supported in part by the National Natural Science Foundation of China under grant
Nos.11405128, 11375239, 11365004, and the Natural Science Foundation of Guangxi province under grant No.2016GXNSFFA380001.
B.A. Li acknowledges the U.S. Department of Energy, Office of Science, under Award
Number DE-SC0013702, the CUSTIPEN (China-U.S. Theory Institute for Physics with Exotic Nuclei)
under the US Department of Energy Grant No. DE-SC0009971, the National Natural Science
Foundation of China under Grant No. 11320101004 and the Texas Advanced Computing Center.

\end{CJK*}

\end{document}